\documentclass[]{aa}  

\usepackage{amsmath}
\usepackage{graphicx}
\usepackage{txfonts}
\usepackage{natbib}

\bibpunct{(}{)}{;}{a}{}{,}

\begin{document}

\title{Hydrogenation at low temperatures does not always lead to saturation: the case of HNCO.}

\authorrunning{Noble et al.~}
\titlerunning{Hydrogenation of HNCO}

\author{J.~A. Noble$^a$, P. Theule$^a$, E. Congiu$^b$, F. Dulieu$^b$, M. Bonnin$^b$, A. Bassas$^a$, F. Duvernay$^a$, G. Danger$^a$, and T. Chiavassa$^a$}

 \institute{a Laboratoire Physique des Interactions Ioniques et Mol\'{e}culaires, UMR 7345-CNRS, Aix-Marseille Universit\'{e}, 13097 Marseille Cedex 20, France\\
   b LERMA-LAMAp, Universit\'{e} de Cergy-Pontoise, Observatoire de Paris, ENS, UPMC, UMR 8112 du CNRS, 5 mail Gay Lussac, 95000 Cergy Pontoise Cedex, France\\
              \email{jennifer.noble@univ-amu.fr}
              }

   \date{Received July 18, 2012; submitted }

\abstract
{It is generally agreed that hydrogenation reactions  dominate chemistry on grain surfaces in cold, dense molecular cores, saturating the molecules present in ice mantles.} 
{We present a study of the low temperature reactivity of solid phase isocyanic acid (HNCO) with hydrogen atoms, with the aim of elucidating its reaction network.}
{Fourier transform infrared spectroscopy and mass spectrometry were employed to follow the evolution of pure HNCO ice during bombardment with H atoms. Both multilayer and monolayer regimes were investigated.} 
{The hydrogenation of HNCO does not produce detectable amounts of formamide (NH$_2$CHO) as the major product. Experiments using deuterium reveal that deuteration of solid HNCO occurs rapidly, probably via cyclic reaction paths regenerating HNCO. Chemical desorption during these reaction cycles leads to loss of HNCO from the surface.}
{It is unlikely that significant quantities of NH$_2$CHO form from HNCO. In dense regions, however, deuteration of HNCO will occur. HNCO and DNCO will be introduced into the gas phase, even at low temperatures, as a result of chemical desorption.}


\maketitle

\section{Introduction}

Isocyanic acid, HNCO, is the simplest molecule containing the four most abundant elements: hydrogen, carbon, nitrogen, and oxygen. It is an important interstellar molecule, with gas phase abundances of around 10$^{-9}$ -- 10$^{-8}$ with respect to H$_2$ in molecular clouds, where it is believed to trace dense, cold gas \citep{Jackson84}. Since its first detection in the Sgr B molecular cloud complex \citep{Snyder72}, HNCO has been detected in multiple environments, including hot cores \citep[e.g.][]{Helmich97}, high mass young stellar objects \citep{Bisschop07}, molecular outflows \citep{Rodriguez10}, comets \citep{Lis97}, and other galaxies \citep{Nguyen91}. It has been shown that HNCO is also a tracer of warm gas, and that its formation origins are likely to be predominantly in icy grain mantles \citep{Bisschop07}.

Isocyanic acid has not yet been observed in the solid state, but it is believed to be responsible for the formation of the abundant cyanate ion OCN$^-$ \citep[e.g.][]{Soifer79,Demyk98,Lowenthal02,vanBroekhuizen05} via reaction of the acidic HNCO with bases such as NH$_3$ \citep{Raunier03a,vanBroekhuizenAA04,Mispelaer12} 
and H$_2$O \citep{Raunier03b,Theule11b}, or by the irradiation of ices with ultraviolet photons \citep{Lacy84} or protons \citep{Moore83}. Moreover, its solid phase chemical reactions have been shown to give rise to isomerisation, with cyanic acid, HOCN, formed thermally in mixtures of HNCO and H$_2$O \citep{Theule11b}. 

The chemical network surrounding HNCO has not been fully studied experimentally. Early theoretical models assumed that HNCO formed only in the gas phase \citep[e.g.][]{Iglesias77}, but more recent studies contend that HNCO forms on grain surfaces via the thermal reaction NH + CO or by the hydrogenation of OCN \citep[e.g.][]{Garrod08,Tideswell10}. Its presence in the gas phase is explained by the subsequent desorption of HNCO from grains, or by the destruction of larger species such as urea, (NH$_2$)$_2$CO.  Abundances of HNCO are enhanced in shocked regions, and thus it is likely that sputtering or additional gas phase formation routes are active in post-shock gases \citep{Zinchenko00}.

The abundance of solid HNCO is predicted to remain relatively low, at $\sim$~10$^{-4}$ with respect to H$_2$O, and thus is not detectable in infrared spectra. If the dominant formation route of HNCO (NH + CO) were efficient \citep{Garrod08}, destruction routes would be required to explain this. As mentioned above, the formation of OCN$^-$ by reaction of HNCO with H$_2$O and NH$_3$ are favourable routes, the reactions having activation barriers of 26~$\pm$~2~kJ\,mol$^{-1}$ (3130~$\pm$~240~K; \citet{Theule11b}) and 0.4~$\pm$~0.1~kJ\,mol$^{-1}$ (48~$\pm$~12~K; \citet{Mispelaer12}), respectively. Irradiation of HNCO with vacuum ultraviolet (VUV) radiation has been shown to produce formaldehyde (H$_2$CO), formamide (NH$_2$CHO), and urea (H$_2$NCONH$_2$) \citep{Raunier04}.

In dense molecular clouds the secondary photon field is weak and hydrogen atoms have a long residence time on grain surfaces because of the low temperature \citep{Tielens82,Amiaud07}. Hydrogen atoms are mobile on the surface at 10~K, and thus hydrogenation reactions dominate low temperature ice chemistry. Experimentally, there still remain many hydrogenation reactions to characterise, although studies have been carried out, particularly on simple molecules. The hydrogenation of CO to form H$_2$CO and CH$_3$OH has been extensively studied \citep[e.g.][]{Hiraoka94,Watanabe02,Fuchs09}, as have the reaction pathways to the formation of H$_2$O from atomic oxygen \citep{Hiraoka98,Dulieu10}, molecular oxygen \citep{Miyauchi08,Ioppolo08,Chaabouni12}, and ozone \citep{Mokrane09}. The formation of C$_2$H$_5$OH, CH$_4$, H$_2$CO, and CH$_3$OH from CH$_3$CHO has been studied by \citet{Bisschop07b}, while the formation of NH$_2$OH by the hydrogenation of NO \citep{Congiu12} and the formation of CH$_2$NH and CH$_3$NH$_2$ from HCN \citep{Theule11a} have recently been demonstrated. 

Given that atomic hydrogen is present at fractional abundances of [H$_\text{I}$]/[H$_2$] $\sim$~10$^{-3}$ in molecular clouds \citep{Li03}, hydrogenation is likely to dominate the destruction pathways of HNCO at low temperatures. Theoretical studies may suggest that the radical intermediates HNCHO or NH$_2$CO form rapidly from HNCO, and the stable molecule formamide forms on further hydrogenation of these intermediate species \citep{Garrod08}, giving overall

\begin{equation}\label{eqn_hnco_h_a}
 HNCO + 2H \rightarrow H_2NCHO.
\end{equation}

This work focuses on the reaction of HNCO with H, which has not yet been examined experimentally. The experiments performed are introduced in \S~\ref{sec-expt}, the results of these experiments are presented and discussed in \S~\ref{sec-results}, while the astrophysical implications of the results are considered in \S~\ref{sec-astro}.

\section{Experimental}\label{sec-expt}

Experiments were performed using two different experimental set-ups: RING, as described in \citet{Theule11b}, and FORMOLISM, as described in \citet{Amiaud06}. 

The RING set-up was used to perform multilayer, bulk ice experiments. Briefly, RING consists of a gold-plated copper surface within a high vacuum chamber (a few 10$^{-9}$~mbar). Molecular species in the form of room temperature gas are dosed onto the gold surface (15 -- 300 K) by spraying via an injection line. The infrared spectra of the molecular solids are recorded by means of Fourier transform reflection absorption infrared spectroscopy (FT-RAIRS) using a MCT detector in a Vertex 70 spectrometer. A typical spectrum has a 1~cm$^{-1}$ resolution and is averaged over a few tens of interferograms. 

The FORMOLISM set-up was used to perform experiments in the monolayer and sub-monolayer regime. The experimental set-up consists of an ultra high vacuum (UHV) chamber (base pressure $\sim$~1~$\times$~10$^{-10}$~mbar), containing a previously oxidised graphite HOPG sample (7 -- 400~K, controlled by a closed-cycle He cryostat). Molecules are dosed onto the surface via two triply differentially pumped beam lines. Desorption of molecules from the surface is monitored using a quadrupole mass spectrometer (QMS, Hiden HAL-3F), positioned directly in front of the surface.

Isocyanic acid was prepared in the gas phase from cyanuric acid (HNCO)$_3$ via thermal decomposition of the commercially available trimer (Aldrich, 98~$\%$) at 650$^\circ$~C under primary vacuum \citep{Raunier03b}. Small quantities of CO$_2$ and traces of CO are always present in the HNCO as a residual of the synthesis method.

In the multilayer experiments presented here, HNCO was dosed onto the surface held at 17~K via an injection line. The HNCO ice was bombarded with H atoms (at approximately 300~K, with a flux of $\sim$~10$^{14}$~cm$^{-2}$s$^{-1}$) produced in a molecular hydrogen plasma generated by a 2.45~GHz microwave discharge. The plasma source and its calibration are fully detailed in \citet{Theule11a}. The hydrogenation was monitored at regular time intervals using IR spectroscopy.
In monolayer and sub-monolayer experiments, HNCO was dosed onto the surface held at 90~K via a molecular beam. This ensured that the deposited HNCO was not mixed with the byproduct CO$_2$. The HNCO was cooled to $\sim$~10~K, then bombarded with H or D atoms (at approximately 300~K, with a flux of 1~$\pm$~0.3~$\times$~10$^{13}$~cm$^{-2}$s$^{-1}$) produced in a hydrogen or deuterium plasma in the second molecular beam. After bombardment, the ices were probed using mass spectrometry during temperature programmed desorption (TPD) experiments.

\section{Results and discussion}\label{sec-results}

\begin{figure}
\centering
\includegraphics[width=0.5\textwidth]{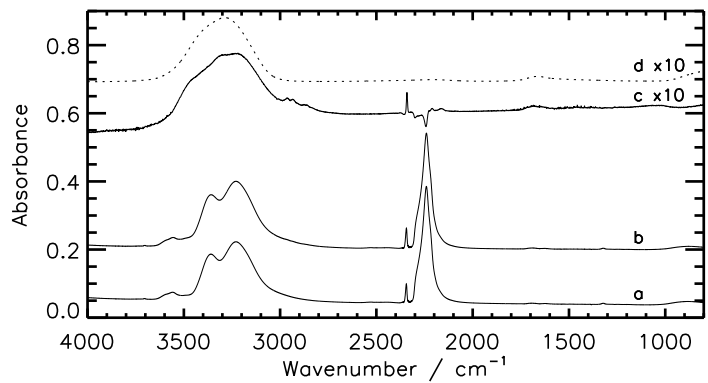}
\caption{ The infrared absorption spectra of pure HNCO bombarded with H. The spectra are as follows: a) pure HNCO deposited at 17~K, b) HNCO bombarded with H atoms for 140 minutes, and c) the difference spectrum of traces a and b. Trace d is a reference spectrum discussed in full in the text.}
\label{figure_hnco}
\end{figure}

\subsection{The multilayer, bulk HNCO regime.}\label{sec-hnco}

\begin{table*}
\caption{Fundamental infrared band positions (cm$^{-1}$), for the species identified in the ice during the bombardment of HNCO$^{a,b,c}$, where $\nu$ represents a stretching vibration, and $\delta$ a bending vibration.}
\label{table_freq}
\begin{center}
\begin{tabular}{lccccccc}
\hline\hline
Species                   &$\nu$N--H              & $\nu$C--H     & $\nu$N=C=O$_{as}$  & $\nu$C$\equiv$O & $\nu$C=O  &$\delta$N--H$_s$ \\ 
\hline
HNCO                     & 3554/3362/3231    &  --        & 2240                  & --                         & --              & --                                                       \\
NH$_2$CHO           & 3313/3169$^{c}$               &  2886$^{c}$              & --                          & --                        & 1685$^{b}$                   & 1385$^{c}$           \\ 
OCN$^{-/\bullet}$       &  --                           & --                   & 2163$^{a}$                    & --                        & --                                     & --        \\
CO                          & --                            & --                   & --                               & 2136$^{b}$             & --                                      &  -- \\
CO$_2$                   & --                            & --                   & --                               & --                          & 2344$^{a}$                          & --          \\
\hline
\end{tabular}
\end{center}
$^a$ Positively identified in the H-bombarded HNCO ice.
$^b$ Tentatively identified in the H-bombarded HNCO ice.
$^c$ Not identified in the H-bombarded HNCO ice. 
\end{table*}

\subsubsection{H bombardment of HNCO}\label{sec_hnco_h}

The spectrum of pure multilayer HNCO deposited at 17~K is presented in Fig.~\ref{figure_hnco}, curve  a. The molecular species HNCO is identified via its characteristic absorption bands, as listed in Table~\ref{table_freq}. The most intense bands are the N--H stretching mode absorptions at 3554, 3362, and 3231~cm$^{-1}$ and the N=C=O asymmetric stretching mode absorption at 2240~cm$^{-1}$. A minor CO$_2$ contamination is identified via the peak at 2344~cm$^{-1}$.

The pure solid HNCO at 17~K was bombarded with H atoms for a total of 140~minutes (corresponding to a dose of $\sim$~8~$\times$~10$^{17}$~cm$^{-2}$). The temperature of the deposition and H bombardment was chosen to be low enough to allow the H atoms to have a relatively long residency time on the surface, while being high enough to allow the H atoms to have a high mobility on the surface, to penetrate the bulk ice as deeply as possible, and to aid in overcoming any potential activation barrier to the hydrogenation of HNCO. The spectrum of the H-bombarded HNCO is presented in Figure~\ref{figure_hnco}, curve b, while the difference spectrum of pure and H-bombarded HNCO is presented in Figure~\ref{figure_hnco}, curve c. The positions of the absorption bands of the observed products are given in Table~\ref{table_freq}.

There are a number of crucial points to note with regard to the difference spectrum of H-bombarded HNCO (Fig.~\ref{figure_hnco}, curve c). First, in the wavelength region 2200 -- 2300~cm$^{-1}$ the peak absorption at 2240~cm$^{-1}$ has diminished, indicative of a decrease of approximately 1 to 2 monolayers (of 580 total monolayers, calculated assuming a band strength of 7.8~$\times$~10$^{-17}$~cm\,molec$^{-1}$ \citep{vanBroekhuizenAA04}). Bands appearing at $\sim$~3300~cm$^{-1}$ and 1680~cm$^{-1}$ are attributed to H$_2$O, and the band at 2342~cm$^{-1}$ is attributed to CO$_2$; both of these species are contaminants introduced via the H plasma, confirmed by blank experiments where the bare gold surface was bombarded. Figure~\ref{figure_hnco}, curve d is a  pure H$_2$O spectrum at 15~K fitted to the 1680~cm$^{-1}$ band of the difference spectrum. Thus fitted, H$_2$O accounts for most of the main feature in the difference spectrum (at $\sim$~3300~cm$^{-1}$) as well as the full 1680~cm$^{-1}$ band. Remaining differences between the spectra of pure and H-bombarded HNCO are relatively minor.

\begin{figure}
\centering
\includegraphics[width=0.5\textwidth]{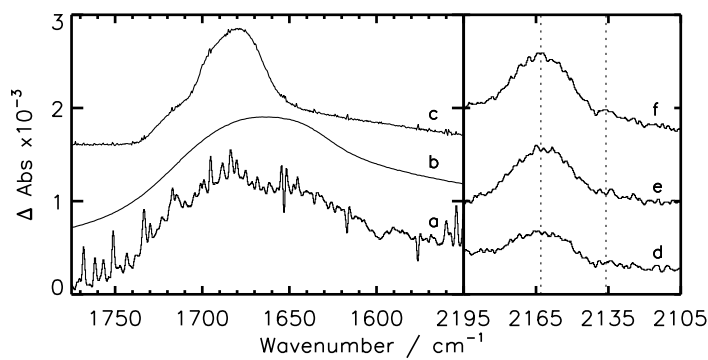}
\caption{The observed products of H + HNCO in the multilayer regime.  In the left  panel, the spectra are as follows: a) the new band at 1680 cm$^{-1}$ (magnification of the difference spectrum in Fig.~\ref{figure_hnco}c), b) a spectrum of pure H$_2$O, scaled to the band at 1680 cm$^{-1}$ (magnification of Figure~\ref{figure_hnco}d), and c) a spectrum of pure NH$_2$CHO, scaled to the band at 1680 cm$^{-1}$. In the right  panel, the spectra are the difference spectra of H-bombarded HNCO and pure HNCO are shown for bombardment times of d) 20 minutes, e) 80 minutes, and f) 140 minutes (magnification of the difference spectrum in Fig.~\ref{figure_hnco}c).}
\label{figure_minor}
\end{figure}

After inspection of the difference spectrum it appears that the expected hydrogenation product, methanamide (formamide, NH$_2$CHO), has not been produced in detectable quantities.
Arguably the simplest potential product, formamide would be formed by the direct hydrogenation of HNCO, as in Eq.~\ref{eqn_hnco_h_a}. The left  panel of Figure~\ref{figure_minor} shows two alternative fits to the 1680~cm$^{-1}$ feature (curve a); curve b is the same scaled H$_2$O spectrum as in Figure~\ref{figure_hnco}, while curve c is a spectrum of pure NH$_2$CHO. The H$_2$O ice feature is a qualitatively better fit to the band formed during the experiment, fully accounting for its width and red wing, while also accounting for the 3300~cm$^{-1}$ feature as discussed above (see Figure~\ref{figure_hnco}, curves c and d). However, traces of NH$_2$CHO may have been formed during H bombardment and its absorption features are masked by the presence of H$_2$O. No other peaks attributable to NH$_2$CHO were observed in the difference spectrum.

One further absorption feature in the difference spectrum can be analysed: a band appearing at 2163~cm$^{-1}$, as illustrated in the right  panel of Figure~\ref{figure_minor} for H bombardment times of 20 (curve d), 80 (curve e), and 140 (curve f) minutes. The slight shoulder, centred at $\sim$~2136~cm$^{-1}$, can be attributed to CO, which is a trace contaminent identified in the H plasma along with H$_2$O and CO$_2$. The  OCN$^-$ ion is known to have an absorption band in the region of 2163~cm$^{-1}$ \citep[e.g.]{vanBroekhuizenAA04} and could form thermally \citep{Theule11b} by reaction with the deposited H$_2$O:

\begin{equation}\label{eqn_hnco_h_d}
 HNCO + H_2O \rightarrow OCN^- + H_3O^+.
\end{equation} 

The counterion H$_3$O$^+$ was not observed in the H-bombarded HNCO ice, but it should be noted that the band strength is very weak \citep{Falk75}, and much lower than that of OCN$^-$ (1.3~$\times$~10$^{-16}$~cm\,ion$^{-1}$ \citep{vanBroekhuizenAA04}), so the absorption features would be vanishingly weak in our ice. If OCN$^-$ is the species responsible for the band centred at 2163~cm$^{-1}$, it is present in very small quantities: approximately 0.1 monolayers account for the peak produced after 140 minutes of H bombardment. However, given the quantity of H$_2$O deposited over the course of the H bombardment, it is not unreasonable that HNCO could react as in Equation~\ref{eqn_hnco_h_d}, as the rate of reaction was found to be 26~kJ\,mol$^{-1}$ \citep{Theule11b}, which could be delivered to the ice surface by room temperature H atoms. 

Another potential, but rather unlikely, explanation for the absorption feature is that the radical species NCO is formed by decomposition of HNCO, probably by H abstraction:

\begin{equation}\label{eqn_hnco_h_e}
 HNCO + H \rightarrow NCO + H_2.
\end{equation}

The spontaneous decomposition of HNCO adsorbed on metal surfaces has been extensively reported in the surface science literature \citep[e.g.][]{Kiss83,Celio97,Jones08}. The NCO radical formed by this decomposition is relatively stable, being observed on the surface at temperatures up to 600~K under UHV conditions \citep{Celio97}. The position of the absorption band associated with the asymmetric stretching mode absorption of NCO is highly dependent on the environment, with a range of around 2145 -- 2305~cm$^{-1}$ \citep[e.g.][]{Nemeth07}. 

One final comment to make about the multilayer reactivity of HNCO is that the VUV irradiation of pure HNCO has previously been shown to produce formaldehyde, formamide, and urea \citep{Raunier04}. In this work, urea was not observed in the IR spectra after H bombardment of HNCO. As the N=C bond must be broken to form urea, this route is unlikely.
The only clear product of H bombardment of HNCO is the species responsible for the absorption band at 2163~cm$^{-1}$; probably OCN$^-$. The other absorption features remain ambiguous.

\subsubsection{H bombardment of NH$_2$CHO}\label{sec_nh2cho}

As the results for the hydrogenation of HNCO were inconclusive, the reactivity of the expected product, formamide, was examined. The hydrogenation of formamide could lead to the production of aminomethanol according to

\begin{equation}\label{eqn_nh2cho_h}
    NH_2CHO + 2H \rightarrow NH_2CH_2OH.
\end{equation}
The molecule NH$_2$CH$_2$OH is the most saturated form of the NCO moiety, so is the logical end-point for the hydrogenation chemistry of HNCO.

\begin{figure}
\centering
\includegraphics[width=.5\textwidth]{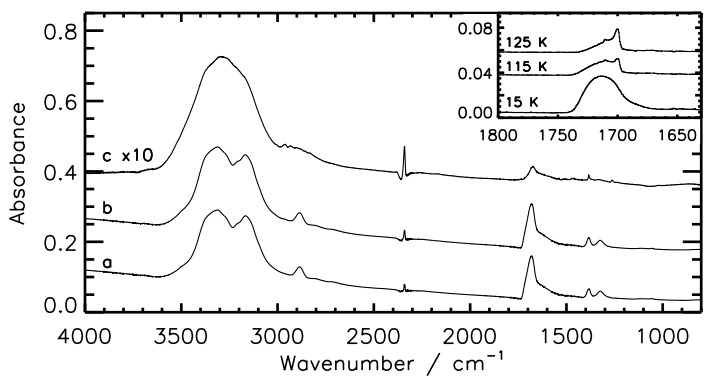}
\caption{Main panel: The infrared absorption spectra of pure NH$_2$CHO bombarded with H atoms. a) pure NH$_2$CHO deposited at 15~K, b) NH$_2$CHO after 190 minutes of H bombardment, c) the difference spectrum of a and b. Inset: Infrared spectra of NH$_2$CHO deposited at 15~K then heated at 2~K\,min$^{-1}$, taken at 15~K, 115~K, and 125~K.}
\label{figure_nh2cho}
\end{figure}

Gas phase NH$_2$CHO was dosed onto the surface held at 15~K, then bombarded with H atoms for 190 minutes using the RING experimental set-up. 
The spectrum of pure NH$_2$CHO at 15~K is presented in Figure~\ref{figure_nh2cho}, curve a, with the H-bombarded sample spectrum as curve b. The difference spectrum of pure and H-bombarded NH$_2$CHO is presented in Figure~\ref{figure_nh2cho}, curve c.

Pure formamide exhibits peaks at 3313, 3169, 2886, 1683, 1385, and 1327 cm$^{-1}$, as seen in Figure~\ref{figure_nh2cho}, curve a. The CO$_2$ molecule was present as a pollutant in the deposited NH$_2$CHO, as shown by the absorption feature in curve a (at 2341~cm$^{-1}$), and in the H plasma, resulting in absorptions in curves b and c. There is no clear decrease observed in the peak of the $\nu$C=O absorption band at 1681~cm$^{-1}$, and thus we conclude that NH$_2$CHO has not reacted. 

Similarly to the bombarded HNCO, the difference spectrum of bombarded NH$_2$CHO is dominated by absorption in the range 3600 -- 3000~cm$^{-1}$. As discussed above, this might have multiple sources, but a large contribution is due to H$_2$O. The four small peaks on the red wing between 2961 and 2859 cm$^{-1}$ are again attributed to contamination of the H plasma due to primary vacuum. Such contamination is only observed because of the very long irradiation times. The peak observed in the difference spectrum at 1676~cm$^{-1}$ is interpreted as further evidence of the deposition of H$_2$O, with a potential contribution from the crystallisation of the NH$_2$CHO sample during H bombardment. We confirmed that, after heating a sample of pure NH$_2$CHO at 2~K\,min$^{-1}$, the crystallisation of NH$_2$CHO results in the shifting of the $\nu$C=O absorption band by 4 -- 5 cm$^{-1}$ towards the blue, as illustrated in the inset of Figure~\ref{figure_nh2cho}. After H bombardment of NH$_2$CHO, no formation of OCN$^-$ was observed. We conclude that formamide is unreactive to H under our experimental conditions, and aminomethanol is not formed via this route.

\subsection{The monolayer HNCO regime.}

Because of the uncertainty in interpretation of our multilayer HNCO data, related to the low level contamination introduced by the H plasma of RING, we decided to perform experiments in the monolayer regime using FORMOLISM. Both experimental set-ups are described in \S~\ref{sec-expt}, but the major advantage of FORMOLISM for this study is a base pressure of $\sim$~10$^{-8}$ mbar in the last stage of the molecular jets, which reduces the level of contamination of the H introduced into the chamber. Working in the monolayer regime allows us to bombard all deposited HNCO with H, unlike the multilayer regime where hydrogen penetrates the first $\sim$~1--3 monolayers, and the signal of the bulk HNCO ice overwhelms that of product molecules in both the Fourier transform infrared (FTIR) and mass spectra. However, the monolayer regime also limits the signal-to-noise ratio of potential products observed using mass spectrometry and, in these experiments, we were unable to observe a clear signal for our HNCO reactant using FTIR spectroscopy. 

\subsubsection{Desorption energy of HNCO}\label{sec_des}

\begin{figure}
\centering
\includegraphics[width=.5\textwidth]{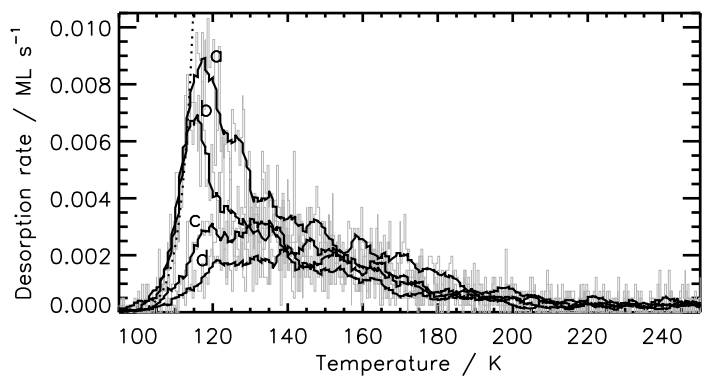}
\caption{Characterisation of the HNCO monolayer deposition. The TPD spectra (m/z 43) for HNCO depositions of a) 1.5 ML, b) 1 ML, c) 0.75 ML, and d) 0.5 ML. The original data are plotted in grey, with a smoothed version overplotted in black. The Polanyi-Wigner fit to the leading edge of the 1.5 ML deposition (see \S~\ref{sec_des}) is plotted as a dotted line.}
\label{fig_desorption}
\end{figure}

Using FORMOLISM, we performed experiments to determine the deposition conditions necessary to deposit 1~ML of HNCO on the graphite surface; HNCO was deposited for a series of fixed times, and TPDs were performed after each deposition. The TPD spectra are shown in Figure~\ref{fig_desorption}. The characteristic monolayer deposition time was determined by visual inspection of the leading edges of these spectra. All HNCO depositions were subsequently quantified by comparison with the identified monolayer deposition (Figure~\ref{fig_desorption}, curve b). The low flux in the molecular beam allows us to deposit very reproducible quantities of molecular species. It takes almost 13 minutes to deposit 1~ML of HNCO, so the deposition uncertainty on coverages of 0.5~ML and 1~ML (used in this work) is vanishingly small and the depositions are highly reproducible. 

The multilayer desorption energy of HNCO was calculated via two different methods.
The rate of desorption by unit surface, $r$, can be expressed by the Polanyi-Wigner equation \citep{Redhead62,Carter62,King75}, where the desorption rate constant $k_{\text{des}}$ is described in terms of an Arrhenius law,

\begin{equation}\label{eqn:polanyi2}
r = -\frac{dN}{dT} = \frac{A}{\beta}\,e^{-E_{\text{des}}/RT}\,N^n,
\end{equation}
where $A$ is the pre-exponential factor, $\beta$~=~$\frac{dT}{dt}$~=~12~K\,min$^{-1}$ is the heating rate, $E_{\text{des}}$ is the energy of desorption of a molecule from the surface (J\,mol$^{-1}$), R is the gas constant (J\,K$^{-1}$\,mol$^{-1}$), $T$ is the temperature of the surface (K), $N$ is the number of adsorbed molecules on the surface (molecules\,cm$^{-2}$), and $n$ is the order of the reaction. The units of $A$ depend on $n$: molecules$^{1-n}$\,cm$^{-2+2n}$\,s$^{-1}$. When analysing the desorption of 1.5~ML of HNCO (Figure~\ref{fig_desorption}, curve a), zeroth order desorption kinetics are assumed, as is standard practise for the multilayer desorption of bulk material, and therefore A has units of molecules\,cm$^{-2}$\,s$^{-1}$. 

By fixing the pre-exponential factor, $A$, at a value of 10$^{28}$~molecules\,cm$^{-2}$\,s$^{-1}$ (assuming that the lattice vibrational frequency of the solid is 10$^{13}$~s$^{-1}$ and the number of molecules in a monolayer is approximately 10$^{15}$~cm$^{-2}$), the desorption energy is calculated as the only free parameter in the fit of Equation~\ref{eqn:polanyi2} to the leading edge of the experimental data \citep{Collings03}. This fit is plotted as a dotted line in Figure~\ref{fig_desorption}. Using this method, the calculated desorption energy was $E_{\text{ads,HNCO}}$~=~32.9~$\pm$~1.7~kJ\,mol$^{-1}$ (3957~$\pm$~204~K).

A second fitting method, proposed by \citet{Hasegawa92}, assumes that the pre-exponential factor in Equation~(\ref{eqn:polanyi2}) is a function of $E_{\text{ads}}$ approximated by

\begin{equation}\label{eqn:harmonic}
  A = N_{\text{ML}} . \nu = N_{\text{ML}} \sqrt{\frac{2N_{\text{ML}}E_{\text{ads}}}{\pi^2M}},
\end{equation}
where $M$ is the mass of the adsorbate molecule, and $N_{\text{ML}}$~$\sim$~10$^{15}$~cm$^{-2}$. The advantage of this method is that the fit requires only one variable, $E_{\text{des}}$, rather than assuming or fitting the pre-exponential factor and fitting $E_{\text{des}}$ \citep{Acharyya07,Noble12}. 
Using the second method, the derived desorption energy was $E_{\text{ads,HNCO}}$~=~31.0~$\pm$~1.6~kJ\,mol$^{-1}$ (3729~$\pm$~192~K).

\begin{figure}
\centering
\includegraphics[width=.5\textwidth]{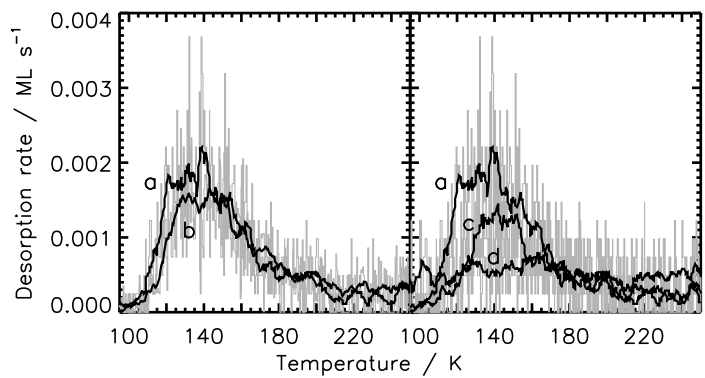}
\caption{H or D bombardment of HNCO. Left  panel: TPDs of a) 0.5~ML HNCO (m/z 43, HNCO), b) 0.5~ML HNCO + 10 minutes H (m/z 43). Right  panel: TPDs of a) 0.5~ML HNCO (m/z 43, HNCO), c) 0.5~ML HNCO + 10 minutes D (m/z 43), d) as for spectrum c, but for m/z 44 (DNCO). The original data are plotted in grey, with a smoothed version overplotted in black. }
\label{figure_halfML}
\end{figure}

\subsubsection{H and D bombardment of HNCO}

A series of experiments were performed on the H or D bombardment of monolayer and sub-monolayer quantities of solid HNCO at 10 -- 15~K. In this temperature range, H atoms do not form a bulk solid because of the rapid self recombination on the surface. A steady state, with an H$_2$ coverage of $\sim$~10~\% \citep{Amiaud07,Kristensen11} due to desorption, is reached within the first ten seconds of H bombardment \citep{Congiu09}. 

Our initial experiments into the monolayer H or D bombardment of HNCO were carried out on short timescales. The results of 10 minutes (\textit{i.e.} 1.7~ML) of H or D bombardment of 0.5~ML HNCO are shown  in Figure~\ref{figure_halfML}. The left panel includes two curves: the TPD spectrum of m/z~43 (HNCO) for 0.5~ML pure HNCO (curve a) and the corresponding spectrum after 10 minutes of H bombardment (curve b). The H-bombarded HNCO sample contains only 0.42~ML HNCO, which represents a 16~\% decrease compared to the pure HNCO sample. No m/z~45 (formamide) was observed to desorb during the TPD experiment, suggesting that no reaction occurred during H bombardment. However, the difference of 0.08~ML HNCO could be, at least partly, due to the low signal-to-noise ratio of the spectra (which is limited by the simultaneous measurement of multiple m/z).

The results of D bombardment, shown in the right panel of Figure~\ref{figure_halfML}, are more revealing. After D bombardment we observe the species DNCO in our ice sample. In Figure~\ref{figure_halfML} the TPD spectrum of m/z~43 (HNCO) for 0.5~ML pure HNCO is again plotted as curve a. The remaining curves, corresponding to 10 minutes of D bombardment (1.7~ML), are m/z~43 (HNCO, curve c) and m/z~44 (DNCO, curve d), respectively. In this experiment, only 0.33~ML of HNCO remain after 10 minutes of D bombardment, representing a decrease of 34~\%. Assuming that the mass spectrometer is equally sensitive to HNCO and DNCO, we calculate the quantity of DNCO on the surface after D bombardment to be $\sim$~0.18~ML. Thus, the total quantity of HNCO and DNCO on the surface represents 0.51~ML, \textit{\emph{i.e.}} equivalent to the quantity of HNCO originally deposited. We will  discuss the implications of the deuteration of HNCO in \S~\ref{sec-mech}. No m/z~47 (NHDCDO, formamide) desorbed during the TPD, therefore we observe a conversion efficiency from HNCO to DNCO of approximately 100~\%.

\subsubsection{The search for formamide}

\begin{figure}
\centering
\includegraphics[width=.5\textwidth]{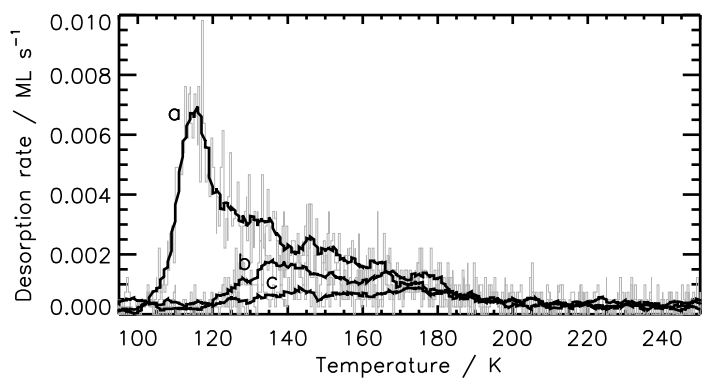}
\caption{Deuteration of HNCO. The spectra are a) reference TPD of 1~ML HNCO (m/z 43), b) TPD after 25~ML (150 minutes) of D bombardment of 1~ML HNCO (m/z 43, HNCO), c) as for spectrum b, but for m/z 44 (DNCO). The original data are plotted in grey, with a smoothed version overplotted in black.}
\label{figure_1ML}
\end{figure}

In order to favour the formation of formamide, we performed an experiment in which we bombarded 1~ML of HNCO with D for 150 minutes (25~ML), comparable to the multilayer bombardment times (see \S~\ref{sec_hnco_h}). The results of this experiment are presented in Figure~\ref{figure_1ML}. A reference TPD of 1~ML pure HNCO (m/z~43) is shown in curve a. The HNCO (m/z~43) and DNCO (m/z~44) on the surface after 25~ML (150 minutes) of D bombardment are shown in curves b and c, accounting for 0.43~ML and 0.22~ML, respectively. Thus, 0.35~ML (35~\%) of the deposited HNCO is unaccounted for after D bombardment. 

\begin{figure}
\centering
\includegraphics[width=.5\textwidth]{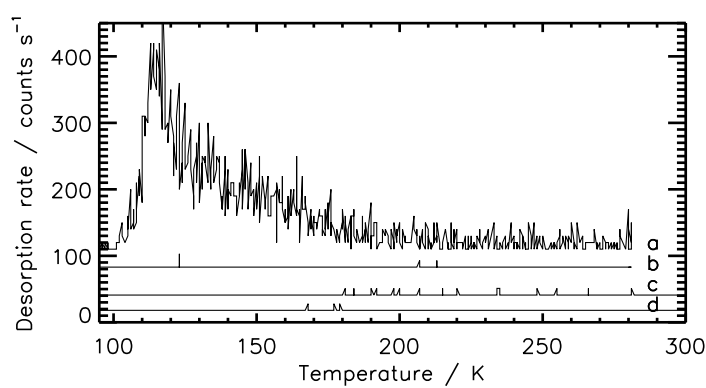}
\caption{Presence of m/z 47 in monolayer experiments. The spectra are a) m/z 43 desorbing during the TPD of 1~ML HNCO, b) the corresponding desorption of m/z 47, c) m/z 47 after 120 minutes of D bombardment of the graphite surface, d) m/z 47 in an HNCO sample bombarded with 25 ML (150 minutes) of D.}
\label{figure_M47}
\end{figure}

In order to determine the origin of the HNCO loss, we searched for the presence of formamide in the D-bombarded HNCO sample. The TPD method is typically sensitive to $\sim$~0.01~ML \citep{Noble11}. In Figure~\ref{figure_M47}, curve a, we show the TPD spectrum of 1~ML pure HNCO (m/z~43) as a reference. All other curves are m/z~47 (NHDCDO). Curve b is from the same experiment as curve a, and indicates that there is no m/z~47 in the deposited HNCO. Curve c is a TPD performed after 120 minutes of D bombardment of the graphite surface, and shows that little or no m/z~47 is present in the atomic beam of D. Curve d is the TPD of the HNCO sample bombarded with D for 150 minutes (see Figure~\ref{figure_1ML}). There is no signal from m/z~47 after D bombardment of HNCO; we thus conclude that formamide is not formed at detectable levels (1~\% of the deposited HNCO) during this experiment.

\subsubsection{Proposed reaction mechanism}\label{sec-mech}

We conclude, based on the evidence from both multilayer and monolayer H and D bombardment of HNCO, that the expected hydrogenation of HNCO to NH$_2$CHO (Equation~\ref{eqn_hnco_h_a}) does not readily occur and thus has a very high barrier. However, it is clear that D bombardment of HNCO produces DNCO and, additionally, that for longer bombardment times there is some loss of HNCO that is not accounted for by this DNCO formation. We provide a list of typical experiments, including the percentage loss of HNCO, in Table~\ref{table_CD}. By comparing the loss of initial HNCO to the doses of H and D, we see that there is a general trend in the data, whereby one molecule of HNCO is lost from the sample for approximately 1 -- 1.5~\% of the incident H and D atoms. Although there was fluctuation in the absolute quantity of HNCO lost from the sample in our experiments, a loss was consistently seen for all H and D exposures greater than $\sim$~2~ML. These losses are likely due to chemical desorption from the surface \citep{Dulieu13}.

\begin{table*}
\caption{The results of experiments carried out on monolayer and sub-monolayer quantities of HNCO.}
\label{table_CD}
\begin{center}
\begin{tabular}{lcccccc}
\hline\hline
Initial HNCO                   & Exposed to     & Temperature & HNCO reduction         & DNCO produced & Loss (to CD) \\ 
(ML)                                &                        & (K)                 & (\%)                            & (\%)                     & (\%)               \\
\hline
0.5                                  & 1.7~ML D        & 10                 & 34 $\pm$ 5                   & 36 $\pm$ 5    & -2 $\pm$ 5       \\
1.0                                  & 10~ML H         & 7                   & 11 $\pm$ 5                   & none               & 11 $\pm$ 5       \\
1.0                                  & 25~ML D         & 15                 & 57 $\pm$ 5                   & 22 $\pm$ 5    & 35 $\pm$ 5       \\
\hline
\end{tabular}
\end{center} 
\end{table*}

Chemical desorption is the stimulated desorption of reactant or product molecules due to an exothermic surface reaction.
Recent experimental studies using FORMOLISM have highlighted the significance of the chemical desorption mechanism in the monolayer regime at low temperatures \citep{Dulieu13,Minissale14,Noble11}. 
It has been shown that up to 90~\% of molecules (on a graphite surface) formed by radical-molecule or radical-radical surface reactions sublimate from the surface after formation because of the inability of the surface to quench the energy released during the reaction. On ice surfaces, more astrophysically relevant for the study of molecules such as HNCO (which will not be present on bare grains), this number is typically much lower.

To account for both the formation of DNCO (after D bombardment) and the loss of HNCO from the surface (after both H and D bombardment), the chemistry is required to proceed via a cyclic pathway, such as the two proposed here:

\begin{subequations}\label{eqn_cycle1}
\begin{align}
 HNCO &\xrightarrow{+ H}{} H_2NCO \xrightarrow{+ H}{} HNCO + H_2, \\
HNCO &\xrightarrow{+ H}{} OCN + H_2 \xrightarrow{+ H}{} HNCO.
\end{align}
\end{subequations}

Cycle~\ref{eqn_cycle1}a corresponds to a classical hydrogenation pathway, \textit{\emph{i.e.}} H addition, but at the second step the branching ratio between the products NH$_2$CHO (Equation~\ref{eqn_hnco_h_a}) and HNCO is fully dominated by the reverse reaction to the initial HNCO population. Cycle~\ref{eqn_cycle1}b corresponds to the abstraction of a hydrogen atom by the incident H atom. The reaction intermediates H$_2$NCO and/or OCN are not expected to be observed in the TPDs as these radicals would either react with incident H atoms at a much faster rate than HNCO or, ultimately, recombine during the heating phase.
It is also possible that, if cycle~\ref{eqn_cycle1}b occurs, the second step could result in the isomerisation of HNCO to HOCN (as previously observed during the heating of mixtures of HNCO and H$_2$O \citep{Theule11b}). However, under current sub-monolayer experimental conditions it is not possible to differentiate between HNCO and HOCN, and no evidence of HOCN was observed in the multilayer experiments.

The results of our D bombardment experiments are critical to explaining the reaction mechanism. At a low dose of D (Table~\ref{table_CD} and Figure~\ref{figure_halfML}), we do not observe any loss of reactants or products from the surface, but deuteration appears to be very efficient. Around 34~\% of the initial HNCO (0.17~ML) is transformed into DNCO. Of the total D atoms incident on the surface, approximately 10~\% have been included in the product. This represents a high efficiency, particularly when we consider that the reaction D + HNCO is in competition with D + D. For comparison, the CO + H system described by \citet{Fuchs09} or \citet{Watanabe02} uses only 1~\% of the incident H atoms in the formation of H$_2$CO and CH$_3$OH. Although we perform our experiments under different conditions, we are able to place the reaction HNCO + H between that of H + H$_2$CO (which is faster) and H + CO (which is slower). It is clear that the barrier to the reaction HNCO + D is not very high, and is probably in the region of 250 -- 700~K. A specific study investigating varying doses and surface temperatures would be required to calculate this value; this represents a large amount of work and is beyond the scope of this paper.

When we extend D bombardment to longer times, we observe a loss of material from the surface in addition to the formation of DNCO (Table~\ref{table_CD} and Figure~\ref{figure_1ML}). For an initial deposition of 1~ML HNCO subjected to 25~ML of D, we recover 43~\% of the HNCO and produce 22~\% DNCO \textit{\emph{i.e.}} a loss of 35~\% of the original material. Assuming that the proportion of D atoms reacting is the same as for short timescales ($\sim$~10~\%), we approximate that HNCO + D occurs 2.5~$\times$ 10$^{15}$\,cm$^{-2}$ times. The observed loss represents 3.5~$\times$ 10$^{14}$\,molec\,cm$^{-2}$, and we can make a rough estimation that the efficiency of chemical desorption is $\sim$~14~\%. Unlike in some previous studies \citep{Dulieu13}, the relative efficiency of the desorption mechanism must be low, as only 0.35~ML is lost from the surface over a bombardment period of 150 minutes. This is equivalent to a desorption rate of $\sim$~4~$\times 10^{-5}$ ML\,s$^{-1}$, and as such any desorbing molecules would not be observable using the mass spectrometer. The result indicates that the reaction driving the chemical desorption mechanism is not highly exothermic.

Each of the two reaction scenarios that we proposed above proceeds via two steps. The first step (H-abstraction or H-addition) limits the kinetics of the overall reaction and is also less likely to provoke chemical desorption since less energy will be released than in the second radical-radical step. Thus most of the desorption should occur at the second reaction step. If the reactivity of both the HNCO and DNCO species was equivalent, we would assume that, with the longer exposures, over 90~\% of the initial HNCO should be transformed, which is not the case. 
The low deuteration yield could be due to the presence of a second reactant (DNCO) and/or steric limitation of the HNCO deposition. However, we propose that the most likely limiting factor is the stabilisation of the HNCO by its formation of dimers or larger polymers, preventing its reaction with H. Experiments performed  on HNCO in a H$_2$O ice, or isolated in a low temperature matricial gas, would reduce the intermolecular bonding and may allow hydrogenation to occur.
Because of the inability to observe desorbing HNCO/DNCO (owing to the low rate of chemical desorption) and the hypothesised intermediates (OCN and/or H$_2$NCO) by mass spectrometry, we are unable to determine whether one reaction cycle dominates the reactivity of HNCO + H. In order to differentiate between the abstraction and addition mechanisms, a full ab initio quantum mechanical calculation treatment including surface effects would be required.

\begin{figure}
\centering
\includegraphics[width=.5\textwidth, trim = 0mm 0mm 0mm 0mm, clip]{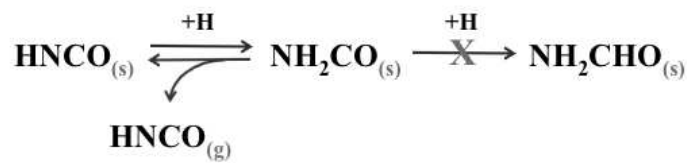}
\caption{Proposed reaction scheme for HNCO with H.}
\label{fig_scheme}
\end{figure}

By comparing our monolayer experiments with those in bulk ice, we confirm that formamide is not produced in detectable quantities after H bombardment of HNCO. It is also plausible that the inconsistency between the loss of HNCO ($\sim$~1 -- 2 ML) and the only confirmed product OCN$^-$/OCN ($\sim$~0.1 ML) in our bulk experiments can be explained by the proposed cyclic reactivity of HNCO combined with chemical desorption. The efficient H/D cyclic substitution suggests that the N-H bond in HNCO is relatively weak, which strengthens the argument for a low level of HNCO decomposition into OCN. We show our proposed overall reaction scheme in Figure~\ref{fig_scheme}.

\section{Astrophysical implications}\label{sec-astro}

Our results contradict theoretical studies which proposed that the initial hydrogenation step to form either HNCHO or NH$_2$CO is very important to HNCO chemistry and that subsequent hydrogenation leads to NH$_2$CHO \citep[e.g.][]{Garrod08,Tideswell10}. However, ab initio calculations determine an activation barrier of 1390~K to the first hydrogenation step \citep{Nguyen96}. A recent experimental study of  the formation of HNCO by addition of the radical NH to CO revealed no formation of NH$_2$CHO after co-deposition of N, H, and CO, despite formation of HNCO \citep{Fedoseev15}.

The NH$_2$CHO molecule has been tentatively identified in ices towards the objects W~33A \citep{Schutte99} and NGC~7538~IRS9 based on comparison with laboratory spectra \citep{Raunier04}, but the identification of molecules in the 6 -- 8~$\mu$m region is complicated by the overlap of multiple absorption bands, and the results of these two studies are not conclusive \citep{Boogert08}. What are the reactions that would drive the chemistry of NH$_2$CHO in the ISM? Although gas phase formation routes to complex molecules exist, it is generally accepted that grain surface chemistry is the dominant formation mechanism for such species \citep{Bisschop07}. The NH$_2$CHO molecule can form by recombination between the radicals NH$_2$ and HCO following energetic processing of ices by, for example, UV photons \citep{Allamandola99,MunozCaro03,vanBroekhuizenAA04}.

It has recently been determined that NH$_2$CHO is the most energetically stable CH$_3$NO isomer that can be formed, and that the amide bond is the most stable bond possible \citep{Lattelais10}. \citet{Jones11} contend that the hydrogenation of HNCO is insignificant as a route to formamide formation because of the low barrier to the reaction HNCO + NH$_3$ \citep{Raunier03a,Mispelaer12}. While it is true that NH$_3$ does react rapidly with HNCO, the flux of hydrogen atoms onto a grain will be superior to the quantity of NH$_3$ molecules in the ice mantle. This is particularly true before formation of NH$_3$ in the mantle. In order to reproduce observed abundances of HNCO and OCN$^-$ in dense molecular clouds, it has been speculated that hydrogenation of HNCO must dominate its destruction mechanisms \citep{Theule11b}. However, our results suggest that hydrogenation is not an efficient process and thus thermal reactions, such as the reaction with NH$_3$ or H$_2$O, should dominate HNCO grain surface chemistry. 

It has already been demonstrated that CH$_3$CHO undergoes hydrogenation after H bombardment, but that it also forms CH$_4$, H$_2$CO, and CH$_3$OH \citep{Bisschop07b}. The hydrogenation product, ethanol, represents approximately 20~\% of the total products formed. The relative strength of the C=O bond prevents hydrogenation being the most prevalent reaction, and the C-C bond breaks to form products that are chemically simpler and of lower mass than the original CH$_3$CHO. Our results confirm that hydrogenation of the C=O bond is not favourable under low temperature H bombardment conditions.
 
Our results show that the OCN moiety is formed in its radical or anionic form during H bombardment.
The OCN$^-$ ion has been positively identified in interstellar ices \citep{Soifer79,Grim87} and its rotational spectrum has been measured \citep{Lattanzi10}, so searches in the gas phase are possible. The neutral OCN \citep{Kawaguchi85} has not yet been identified in the ISM; it has, however, long been predicted in dense clouds \citep{Prasad78}. If OCN/OCN$^-$ can be formed by the reaction of H with HNCO, this could help explain the destruction of HNCO, while confirming that formamide formation does not occur via this route. Additionally, in dense regions such as pre-stellar cores, H/D exchange is potentially the dominant destruction mechanism for HNCO \citep{Roberts03}. 

We have also shown that formamide does not react with hydrogen to produce aminomethanol, the logical saturated end-point for HNCO hydrogenation. 
Aminomethanol has, however, already been shown to form under interstellar conditions by the purely thermal reaction between H$_2$CO and NH$_3$, with an activation barrier of 4.5~$\pm$~0.5~kJ\,mol$^{-1}$ (541~$\pm$~60~K, \citet{Bossa09}). Theoretical studies suggest that, in the presence of acids, aminomethanol can undergo spontaneous dehydration to form methylenimine (CH$_2$=NH) and water \citep{Walch01}; experiments have shown that aminomethanol yields hexamethylenetetramine (HMT) from the polymerisation of the dehydration product methylenimine \citep{Bernstein95,Duvernay11}. Although it has not yet been observed in the interstellar medium, in the laboratory HMT has a sublimation temperature of $\sim$~553~K \citep{Bernstein95}, so it is thought to be present as part of an organic residue on comets.
The formation routes to aminomethanol and hydroxyacetonitrile (HOCH$_2$CN) are competitive when CN$^-$ is included in the initial ice \citep{Danger12}. These destruction routes and competitive reactions, combined with our conclusion that aminomethanol does not form by the hydrogenation of formamide, might result in low interstellar abundances of aminomethanol, although it has not yet been observed.

\begin{acknowledgements}
This work has been funded by the French national program Physique Chimie du Milieu Interstellaire (PCMI) and the Centre National d'Etudes Spatiales (CNES). J.~A.~N. is a Royal Commission for the Exhibition of 1851 Research Fellow.
\end{acknowledgements}

{}

\end{document}